\documentclass[11pt,twoside,twocolumn]{article}

\usepackage{amsmath,amsfonts,amssymb} 
\usepackage[USenglish]{babel}

\newcommand{\Z}{{\mathbb Z}}

\newtheorem{theorem}{Theorem}
\newtheorem{definition}{Definition}
\newtheorem{lemma}{Lemma}

\newenvironment{proof}
  {\noindent \textbf{Proof:}}
  {\qed}
\newenvironment{proof2}
  {\noindent \textbf{Proof of theorem:}}
  {\qed}

\def\qed{\ifmmode\square\else{\unskip\nobreak\hfil
\penalty50\hskip1em\null\nobreak\hfil$\square$
\parfillskip=0pt\finalhyphendemerits=0\endgraf}\fi}

\newcommand{\ket}[1]{|#1\rangle}

\newcommand{\e}{\mathrm{e}}

\newcommand{\ignore}[1]{}
\newcommand{\mattwoc}[4]{\left[
        \begin{array}{cc}{#1}&{#2}\\{#3}&{#4}\end{array}\right]}

\title{\textbf{Quantum clock synchronization with one qubit}}
\author{Chris Harrelson \\
UC Berkeley \\
chrishtr@cs.berkeley.edu
\and
Iordanis Kerenidis \\
UC Berkeley\\
jkeren@cs.berkeley.edu
}

\begin{document}
\maketitle

\nocite{*}
\bibliographystyle{alpha}

\begin{abstract}
The clock synchronization problem is to determine the time difference
$T$ between two spatially separated parties. We improve on I. Chuang's 
quantum clock synchronization algorithm and
show that it is possible to obtain $T$ to $n$ bits of accuracy while
communicating only one qubit in one direction and using an $O(2^n)$ frequency
range. We also prove a quantum lower bound of $\Omega(2^n)$ for the product
of the transmitted qubits and the range of frequencies,
thus showing that our algorithm is optimal. 
\end{abstract}

\section{Introduction}
Clock synchronization is a well studied problem with many practical
and scientific applications.  In the special theory
of relativity there are two standard methods for synchronizing a pair
of spatially separated clocks, Einstein Synchronization
\cite{Einstein1905} and Eddington's Slow Clock Transport
\cite{Eddington1924}.

Recently, two new quantum protocols have been proposed for
synchronizing remote clocks.  The first one uses prior quantum bit
entanglement between the two parties and was proposed by \emph{Jozsa
et al} {\cite{Jozsa}}. This protocol is based on the assumption that
the entanglement can be achieved without any relative phase error.
However, the validity of this assumption has been discussed and
questioned in a number of papers \cite{reply,lorentz,genovese}.  Once
and if this entanglement can be obtained, their algorithm determines
the time difference $T$ between the clocks by essentially monitoring
the oscillation of a function $f(T)\sim \cos(\omega T)$, and thus
requires $O(2^{2n})$ shared singlets.

The second protocol was proposed by {\em I. Chuang} {\cite{Chuang}},
and obtains $T$ to $n$ bits of accuracy by communicating only $O(n)$
qubits and using an $O(2^n)$ range of frequencies.
After communicating the bits according to his protocol, they are in
the state corresponding to the Fourier Transform (over $\Z_{2^n}$) of
the state $\ket{\omega T}$, for some fixed and known $\omega$.  As a
result, one can apply an inverse Fourier Transform and subsequently
measure the value of $\omega T$ and hence $T$.

In this paper, we improve significantly on Chuang's result by
presenting an algorithm that is able to calculate
$T$ to $n$ bits of accuracy while communicating only one qubit 
in one direction and
using an $O(2^n)$ range of frequencies. Further, we prove that,
under our computational model, the product of the
frequency range and the number of transmitted qubits must be
$\Omega(2^n)$, and conclude that our algorithm is optimal in this model.

\section{The computational model}

In our protocol Alice sends a photon $\ket{\psi}$ to Bob with some
tick rate $\omega$. The state
of the received photon is $\e^{i\omega t Z}\ket{\psi}$, where $t$ is
the time the photon spent in transit and $Z$ is the Pauli matrix
\[ Z =\mattwoc{1}{0}{0}{-1}.\] 
Even though we only use one-way communication from Alice to Bob, for
the purposes of proving computational lower bounds we assume an even
stronger model, where the two parties can exchange photons back and
forth. The information they get about the time difference $T$ between the
two clocks comes from a phase change in the state of the qubit, which
depends on $T$ and the tick rate $\omega$. Let's now define this procedure
and see how we can actually implement it. The input to this procedure will be a quantum register which
holds the tick rate $k$ and a qubit $\ket{\psi}$. The output is a state
that has a phase which depends on $T$ and $k$.
\begin{definition} Let \textsc{tqh} be a black box quantum procedure
defined by the equation
\[ \textsc{tqh}(\ket{k}\ket{\psi}) = 
\ket{k}e^{2\pi ik \omega_0 TZ}\ket{\psi}\]
where $T$ is the time difference between the two parties and $\omega_0$
is a known base tick rate.
\end{definition}

This is a very reasonable and powerful model, since we know that all
the information one can get about the time difference via such photon
communications is in the form of a relative phase change.  Here the
first register handles the tick rate of the photon to be transmitted
and the second register is the photon that Alice communicates to Bob
(or Bob to Alice).

The implementation of this black box is based on the ticking qubit 
handshake protocol (TQH) described in I. Chuang's
paper \cite{Chuang}. 
Suppose Alice wants to create the state $e^{2\pi ik \omega_0 TZ}\ket{\psi}$.
She first sends the qubit $\ket{\psi}$ to Bob with ticking rate 
$(-2 \pi k \omega_0)$.
Along a classical channel she also tells him her time $t_{A}$ at the
moment of the quantum communication.  Bob receives at time $t_{B}$
(according to Bob's clock) a quantum state $e^{-2 \pi i k \omega_0
t_{tr}Z}\ket{\psi}$, where $t_{tr}$ is the time the qubit spent in
transit. Finally, Bob applies a phase change $e^{2 \pi i k \omega_0
(t_B-t_A) Z}$ and thus the final state of the qubit is $e^{2 \pi i k \omega_0 
(t_B-t_A-t_{tr}) Z}\ket{\psi} = e^{2 \pi i k \omega_0 T Z}\ket{\psi}$.  

\section{An optimal Quantum Algorithm}

We are going to describe a protocol for synchronizing two remote
clocks by communicating one photon. In this algorithm, Alice starts by
preparing a register $R$ of $n$ qubits in a certain superposition. 
Then she  sends a photon to Bob with the superposition of tick rates
specified in $R$. Bob measures the received photon and 
Alice  obtains $T$ to $n$ bits of accuracy by processing a phase
estimation on  $R$.

In more detail, 
\begin{enumerate}
\item Alice starts with a register of $n$ qubits initialized to $\ket{0}$, 
and after applying a Fourier Transform to them she obtains
\[\frac{1}{\sqrt{2^{n}}}\sum_{k\in\Z_{2^n}}\ket{k}.\]
She also prepares a photon with polarization state $\frac{1}{\sqrt{2}}(\ket{0}+\ket{1})$.
\item Alice now transmits the prepared photon with the tick rate
described by her first register. If the photon had a definite tick
rate $k$ and polarization state $\ket{\psi}$ the final
 state would be $e^{2 \pi i k \omega_0 T Z}\ket{\psi}$.  Since
the register described in step 1 is in a superposition of tick rates,
the outcome will be in a superposition of states
\[ \frac{1}{\sqrt{2^{n+1}}} \sum_{k \in \Z_{2^n}} \ket{k} e^{2\pi i (k\cdot\omega_0) T} \ket{\psi}  \]
\item Bob measures the received photon.  Assuming without loss of generality
that the outcome is $\ket{0}$, Alice's register $R$ becomes
\[ \frac{1}{\sqrt{2^n}} \sum_{k \in \Z_{2^n}} e^{2\pi i k \omega_0 T} \ket{k}.\]
\item Alice then applies an inverse Fourier Transform, obtaining the state
$\ket{\omega_0 T}$ in $R$.
\end{enumerate}

It is easy to see that this algorithm is an application of the general 
procedure known as {\it phase estimation}. In this procedure, we assume
a unitary operator $U$ with an eigenvector $\ket{u}$ and eigenvalue $e^{2\pi i \phi}$.
The goal is to estimate $\phi$ to $n$ bits of accuracy. 
To perform the estimation we start 
with two registers, the first one in a uniform superposition over all states in
$\Z_{2^n}$ and the second one in the state $\ket{u}$. Then we apply
the unitary operation $U$ to the second register $j$ times, where
$\ket{j}$ is the content of the first register. 
By analyzing the performance of this procedure it can be seen that
our algorithm obtains $T$ to $n$ bits of accuracy with
constant probability $4/\pi^2$. We can boost the probability of
success to $1-\delta$ by increasing the size of the first register to
$n+\log(2+\frac{1}{2\delta})$.  Further analysis can be found in
\cite{chuang-quantum-book}, page 221.

\section{A lower bound on frequency range $\times$ number of qubits}

In this section we will prove a lower bound on the product of the
range of tick rates (frequencies) we use and the number of qubits
 we communicate.

\begin{theorem}
Any quantum algorithm which determines $T$
to $n$ bits of accuracy, using a range $F$ of frequencies and
communicating $Q$ qubits between the two parties, must have that $F\cdot
Q=\Omega(2^n)$.
\end{theorem}
In order to prove this theorem we are going to use the following lemma:
\begin{lemma}
If a quantum algorithm in the \textsc{tqh} model makes only queries to
the black box with a single tick rate $\omega$, then it must
make a total number of $\Omega(2^n)$ queries in order to obtain $T$ to
$n$ digits of accuracy.
\end{lemma}
\begin{proof}
By making a query to the black box, the input $\ket{\psi}\equiv\frac{1}{\sqrt{2}}(\ket{0}+\ket{1})$
will become  $\e^{2\pi i\omega Z T}\ket{\psi}$;
after applying a Hadamard transform we obtain the quantum state
\[ \cos(2 \pi \omega T)\ket{0}+\sin(2\pi \omega T)\ket{1} .\]
From this we see that the problem of determining $T$ is equivalent to
estimating the amplitude of $\ket{0}$ (or $\ket{1}$). The problem of
estimating the amplitude of a quantum state, which is equivalent to
the problem of counting the number of solutions to a quantum problem,
is well-studied \cite{Hoyer,Nayak}.  In \cite{Nayak} they prove that
$\Omega(\sqrt{N/\Delta}+\sqrt{t(N-t)}/\Delta)$ queries are required
for a $\Delta$-approximate count, where $t$ is the number of
solutions, $N$ is the set of possible inputs and $\Delta$ defines
the closeness of the approximation.

If we use this lower bound for the case of amplitude
estimation, we get a lower bound of
$\Omega(\sqrt{N/\Delta}+N\sqrt{a(1-a)}/\Delta)$, since the amplitude
is $a=t/N$.  In our case, $a$ can take any value in $(0,1)$,
$\Delta$ must be less than $1$ and $N = 2^n$, so we obtain the lower bound of
$\Omega(2^n)$ qubits.\\
\end{proof}
Now we are ready to prove Theorem 1.\\\\
\begin{proof2}
Suppose we are able to query the black box with frequencies in the range 
$[\omega,F\omega]$. We claim that this black box can be simulated by a black
box with only one tick rate $\omega$ at the cost of replacing each query 
with at most $F$ queries. This can be done since one query to the 
$[\omega,F\omega]$ black box with tick rate $(k\omega), 1\leq k\leq F$ 
is equivalent to $k$ consecutive queries with 
tick rate $\omega$, using the output of one query as the input to the next.
Notice that a superposition of queries to the $[\omega,F\omega]$ black box
does not pose any challenge to the simulation, since we can also query
the one-tick rate black box in a superposition of times.  For such an input,
the number of queries is defined to be the maximum over all states of the
superpositions.

Since in all cases we query the black box at most $F$ times, this
means that the one-tick rate version will run with at most $F \cdot Q$
queries.  Now, in Lemma 1 we have already proved that when we use only
one tick rate, we need to communicate at least $\Omega(2^n)$ qubits,
and therefore $F\cdot Q=\Omega(2^n)$.
\end{proof2}

\bibliography{}

\end{document}